\documentclass[aps, pra, superscriptaddress, amsmath, amssymb, twocolumn, amsfonts, floatfix, longbibliography]{revtex4-2}
\usepackage{mathtools}
\usepackage{braket}
\usepackage[dvipsnames]{xcolor}
\usepackage{float}
\usepackage{subfigure}
\usepackage{upgreek}
\usepackage{pgffor}
\usepackage{float}
\usepackage{silence}
\usepackage[colorlinks=true,linktoc=page,linkcolor=blue,citecolor=magenta,urlcolor=Fuchsia]{hyperref}
\usepackage[normalem]{ulem}
\usepackage{sidecap,tikz}
\usepackage{orcidlink}
\usepackage{makecell}
\usepackage{mathrsfs}

\newcommand{\beq}{\begin{equation}}
\newcommand{\eeq}{\end{equation}}
\newcommand{\bea}{\begin{eqnarray}}
\newcommand{\eea}{\end{eqnarray}}

\mathchardef\mhyphen="2D % Define a "math hyphen"

\newcommand{\non}{\nonumber}

\begin{document}

% \title{Nonlinear Magnetic Response as a Probe of Spin-Resolved Quantum Geometry}

% \title{Spin-Rotation Quantum Geometry and Nonlinear Magnetic Response in Two-Dimensional Systems}

\title{Intrinsic Nonlinear Gyrotropic Magnetic Effect Governed by Spin-Rotation Quantum Geometry}

\author{Neelanjan Chakraborti}
% \email{neelanjanc23@iitk.ac.in}
\affiliation{Department of Physics, Indian Institute of Technology, Kanpur 208016, India}
\author{{Snehasish Nandy}}
\thanks{Jointly supervised this work}
\email{snehasish@phy.nits.ac.in}
\affiliation{Department of Physics, National Institute of Technology Silchar, Assam 788010, India}
\author{{Sudeep Kumar Ghosh}\,\orcidlink{0000-0002-3646-0629}}
\thanks{Jointly supervised this work}
\email{skghosh@iitk.ac.in}
\affiliation{Department of Physics, Indian Institute of Technology, Kanpur 208016, India}

\date{\today}

\begin{abstract}
Nonlinear magnetic response driven by time-periodic magnetic fields offers a distinct route to probe spin-resolved quantum geometry beyond conventional electric-field-driven nonlinear effects. While linear magnetic responses depend on the Zeeman quantum geometric tensor, the influence of generalized spin-rotation quantum geometries on nonlinear responses has not been established. Here, we develop a microscopic quantum-kinetic framework to elucidate how the Zeeman and spin-rotation quantum geometric tensors govern nonlinear gyrotropic magnetic transport in two-dimensional systems. We derive second-order gyrotropic magnetic currents and reveal a distinct geometric separation: the off-diagonal sector is controlled by the Zeeman symplectic and metric connections, whereas the diagonal sector is dictated by the spin-rotation quantum metric and Berry curvature. This identifies the spin-rotation quantum geometric tensor as a fundamental geometric quantity unique to the nonlinear regime. Applying our theory to massless Dirac fermions, hexagonally warped topological insulator surface states, tilted massive Dirac fermions, and parity-time symmetric CuMnAs, we demonstrate how specific symmetries selectively activate conduction and displacement channels. Our findings link spin-resolved quantum geometry to nonlinear magnetic transport, offering design principles for engineering tailored nonlinear magnetic responses in optoelectronic and spintronic devices.

\end{abstract}

\maketitle

\section{Introduction}

Nonlinear transport phenomena have emerged as a powerful probe of the hidden geometric and topological structure of Bloch bands in crystalline materials~\cite{Suarez2025,Jiang_2025_rpp}. Effects such as second-harmonic generation, bulk photovoltaic currents, and the nonlinear Hall effect are highly sensitive to broken spatial and time-reversal symmetries, directly reflecting microscopic quantum geometric quantities like the Berry curvature and quantum metric~\cite{Du_2021,Ortix2021,Bandyopadhyay2024,Sipe2000,Morimoto2016}. These responses are particularly pronounced in two-dimensional materials, where reduced dimensionality, strong spin–orbit coupling, and valley degrees of freedom enhance the signal magnitude~\cite{Kamal_2023_prb,kamal_2022_prl}. Consequently, 2D systems provide an ideal platform to access the quantum geometric tensor (QGT), whose components govern a wide array of nonlinear charge, spin, and optical responses~\cite{Suarez2025,Jiang_2025_rpp}. Recent advances have demonstrated that beyond the Berry-curvature dipole, the quantum-metric dipole and higher-order multipoles of QGT can dominate transport in materials with subtle symmetry constraints~\cite{Du_2021,Ortix2021}. Experimental observations in transition-metal dichalcogenides, polar semiconductors, and few-layer topological insulators have confirmed exceptionally large nonlinear Hall and shift-current responses that trace this underlying geometry~\cite{Bandyopadhyay2024,Liu2024,Suarez2025}. Furthermore, engineered systems like twisted bilayer graphene and altermagnetic heterostructures are emerging as tunable platforms to selectively control these competing geometric contributions. Together, this progress highlights nonlinear transport as a versatile tool for exploring spin-resolved quantum geometry~\cite{Suarez2025,Jiang_2025_rpp,Liu2024,Du_2021}.

While standard quantum geometry arises from momentum-space translations of Bloch states, a complete description of spin-orbit-coupled bands requires generalizing this framework to include local spin rotations. This extension introduces the Zeeman quantum geometric tensor (ZQGT) and the spin-rotation quantum geometric tensor (SRQGT), which couple position-operator matrix elements with interband spin matrix elements~\cite{Xiang_2025_PRL,Xiao_2010,xiang2025_em,Xiang2025_spin,Jia_2025}. These novel contributions obey distinct symmetry rules: for instance, the Zeeman Berry curvature can survive under time-reversal symmetry (TRS), whereas the Zeeman quantum metric is odd—behavior that allows for both symmetric and antisymmetric components in Berry-curvature and quantum metric, forbidden in the conventional QGT. This enriched structure enables novel transport responses in spin–orbit-coupled and symmetry-broken materials~\cite{Xiang_2025_PRL,Zhong_2016,ghorai2026,Chakraborti2025,chakraborti_2025}. However, a critical gap remains: while Zeeman-driven linear effects, such as intrinsic gyrotropic magnetic currents~\cite{Xiang_2025_PRL,Zhong_2016,Nisarga2025}, successfully probe the ZQGT, the SRQGT remains inactive in the linear regime. Consequently, the influence of spin-resolved quantum geometry on nonlinear transport and specifically the characterization of responses driven by the SRQGT, has yet to be systematically explored.

A natural physical context to probe these spin-rotation induced quantum geometries is natural optical activity, or natural gyrotropy which is the intrinsic ability of an inversion-asymmetric medium to distinguish between right- and left-circularly polarized light. In the low-frequency limit, the current density $\mathbf{j}$ induced in a clean metal by a slowly varying magnetic field $\mathbf{B}$ captures this physics via the dynamical chiral magnetic effect (or gyrotropic magnetic effect), which can be expanded as:
$j_i = \alpha_{ij} B_j + \gamma_{ibc} B_b B_c + \cdots$.
Here, the linear order term, known as the linear gyrotropic current, arises from intrinsic orbital and spin magnetic moments as well as the Zeeman quantum geometry. The second-order term , however, is dictated by the spin-rotational quantum geometry, capturing the nonlinear interplay between spin texture and crystal momentum. Physically, the imaginary parts of these responses manifest as circular dichroism (differential absorption), while their real parts correspond to rotatory power (optical rotation). This hierarchy identifies the nonlinear gyrotropic current as the essential observable for isolating the SRQGT.

In this context, we develop a microscopic quantum-kinetic theory for second-order nonlinear magnetic responses induced by time-dependent magnetic fields in two-dimensional systems. We demonstrate that the second-order gyrotropic magnetic current is governed by the SRQGT and ZQGT, along with their associated symplectic connections, which generate distinct contributions in both conduction and displacement channels. Applying this framework to representative model systems with different symmetries, we elucidate how specific broken symmetries selectively activate different transport channels. We identify the specific symmetry-allowed tensor components responsible for these effects and show that SRQGT contributions, which are silent in linear response, form the core of the nonlinear magnetic transport. Collectively, our results establish a unified geometric framework for understanding and engineering nonlinear magnetotransport in low-dimensional quantum materials.

% %------------------------------------------------------------
% \begin{figure}[t]
% \centering
% \includegraphics[width=\columnwidth]{Schematic_mod.pdf}
% \caption{Nonlinear gyrotropic magnetic transport arises from the application of two perpendicular (or, more generally, in-plane) magnetic fields. These fields activate distinct mechanisms leading to nonlinear gyrotropic magnetoconductivities, studied here across various symmetry-breaking phases of the Dirac model.}
% \label{fig:schematic}
% \end{figure}
% %------------------------------------------------------------

\section{Generalized Quantum Geometry of Bloch states}
Quantum geometry, which measures the infinitesimal distance between neighboring quantum states in Hilbert space, has become a key concept for understanding diverse phenomena in quantum materials. It is described by the quantum geometric tensor (QGT), or Fubini–Study metric, whose real and imaginary parts respectively define the quantum metric and the Berry curvature. In its conventional form, this distance depends solely on crystal momentum, neglecting possible contributions from spin degrees of freedom.
When both momentum translation and spin rotation are included, the infinitesimal quantum distance between neighboring cell-periodic Bloch states $\ket{u^{\zeta}_{m\mathbf{k}}}$ takes the form~\cite{Xiang_2025_PRL}
%------------------------------------------------------------
\begin{align}
    ds^2 &= \left\lVert \mathscr{U}_{d\theta} \mathscr{U}_{d\mathbf{k}} \ket{ u_{m\mathbf{k}}^\zeta} - \ket {u_{m\mathbf{k}}^\zeta} \right\rVert^2 \nonumber \\
     &= \sum_{p \neq m} r_{mp}^{a} r_{pm}^{b} \, dk_a \, dk_b 
    + \tfrac{1}{4} \sum_{m} \sigma_{mp}^{a}\sigma_{pm}^{b} \, d\theta_a \, d\theta_b \nonumber \\ 
   &\quad +\sum_{p \neq m} r_{mp}^{a}\sigma_{pm}^{b} \, d\theta_b \, dk_a, \nonumber \\
    &= \sum_{p \neq m} \mathscr{G}_{mp}^{ab} \, dk_a \, dk_b 
    + \tfrac{1}{4} \sum_{m} \mathscr{S}_{pm}^{ab} \, d\theta_a \, d\theta_b \nonumber \\ 
   &\quad +\tfrac{1}{2} \sum_{p \neq m} \left(\mathscr{Z}_{mp}^{ba} + \mathscr{Z}_{pm}^{ba} \right) \, d\theta_b \, dk_a ,
    \label{eq:qgt}
\end{align}
%-----------------------------------------------------
where $a,b$ are spatial indices and $\zeta$ labels spin. Here, $\mathscr{U}_{d\mathbf{k}} = e^{-i d\mathbf{k} \cdot \hat{\mathbf{r}}}$ generates infinitesimal momentum translations and $\mathscr{U}_{d\theta} = e^{-i d\theta \cdot \hat{\boldsymbol{\sigma}}/2}$ infinitesimal spin rotations via the Pauli matrices $\hat{\boldsymbol{\sigma}}$ (with $\hbar=1$). The tensors $\mathscr{G}$, $\mathscr{S}$, and $\mathscr{Z}$ respectively denote the conventional QGT (CQGT), spin-rotation QGT (SRQGT), and Zeeman QGT (ZQGT), capturing quantum geometry due to momentum-space translations, spin rotations, and their interplay. A unified notation follows from defining $z_{mp}^{ab}=x_{mp}^a y_{pm}^b$, whose real and imaginary parts are
%------------------------------------------------------------
\begin{align}
z_{mp}^{ab,R} &= \tfrac{1}{2}\left(x_{mp}^a y_{pm}^b + x_{pm}^a y_{mp}^b\right), \nonumber \\
z_{mp}^{ab,I} &= i\left(x_{mp}^a y_{pm}^b - x_{pm}^a y_{mp}^b\right).
\label{eq:UnifiedQGT}
\end{align}
%------------------------------------------------------------
% \begin{align}
% z_{mp}^{ab,R} = \tfrac{1}{2}(x_{mp}^a y_{pm}^b + x_{pm}^a y_{mp}^b), \quad
% z_{mp}^{ab,I} = i(x_{mp}^a y_{pm}^b - x_{pm}^a y_{mp}^b).
% \label{eq:UnifiedQGT}
% \end{align}
For $x=y=r$, one recovers the CQGT, where $z^{R}\equiv\mathcal{Q}$ (conventional quantum metric (CQM)) and $z^{I}\equiv\Omega$ (conventional Berry curvature (CBC))~\cite{Xiao_2010,Jiang_2025,Gao_2014,Du_2021}. For $x=y=\sigma$, one obtains the SRQGT, with $z^{R}\equiv\mathscr{R}$ (spin-rotation quantum metric (SRQM)) and $z^{I}\equiv\Lambda$ (spin-rotation Berry curvature (SRBC))~\cite{Jia_2025}. The mixed case $x=r$, $y=\sigma$ defines the ZQGT, whose real and imaginary parts correspond to the Zeeman quantum metric (ZQM, $\mathcal{F}$) and Zeeman Berry curvature (ZBC, $\Gamma$). While both the conventional and spin-rotation Berry curvatures are purely antisymmetric, the Zeeman Berry curvature can possess both symmetric and antisymmetric parts; the antisymmetric component can be written in vector form as $\Gamma^{A}_{n} = \nabla_{\mathbf{k}} \times \boldsymbol{\sigma}_{n}/2$. Likewise, the Zeeman quantum metric can contain an antisymmetric contribution absent in the conventional and spin-rotation counterparts. The transformation properties of these generalized quantum geometric quantities under inversion ($\hat{\mathcal{P}}$), time-reversal ($\hat{\mathcal{T}}$), and combined $\hat{\mathcal{P}}\hat{\mathcal{T}}$ symmetries are summarized in Table~\ref{tab:Symmetry_QGT}.

%%%%%%%%%%-------------------------------------------------------------
\begin{table}[b!]
\centering
\renewcommand{\arraystretch}{1.5} % increase row height
\setlength{\tabcolsep}{18pt} % increase column width
\begin{tabular}{|c|c|c|c|}
\hline
\textbf{Quantities} & $\boldsymbol{\hat{\mathcal{P}}}$ & $\boldsymbol{\hat{\mathcal{T}}}$ & $\boldsymbol{\hat{\mathcal{P}}\hat{\mathcal{T}}}$ \\
\hline
CBC ($\Omega^{ab}_{mp}$) & $+$ & $-$ & $-$ \\
\hline
SRBC ($\Lambda^{ab}_{mp}$) & $+$ & $-$ & $-$ \\
\hline
ZBC ($\Gamma^{ab}_{mp}$) & $-$ & $+$ & $-$ \\
\hline
CQM ($\mathcal{Q}^{ab}_{mp}$) & $+$ & $+$ & $+$ \\
\hline
SRQM ($\mathscr{R}^{ab}_{mp}$) & $+$ & $+$ & $+$ \\
\hline
ZQM ($\mathcal{F}^{ab}_{mp}$) & $-$ & $-$ & $+$ \\
\hline
ZMC ($\mathscr{L}_{\text{pm}}$) & $-$ & $+$ & $-$ \\
\hline
ZSC ($\tilde{\mathscr{L}}_{\text{pm}}$) & $-$ & $-$ & $+$ \\
\hline
\end{tabular}
\caption{Symmetry properties of the generalized quantum geometric quantities under inversion ($\hat{\mathcal{P}}$), time reversal ($\hat{\mathcal{T}}$), and their combination ($\mathcal{PT}$). `$+$' indicates that the quantity is even under the operation, while `$-$' indicates that it is odd. Abbreviations:- Berry curvatures: i) CBC: conventional Berry curvature, ii) SRBC: spin-rotation Berry curvature, and iii) ZBC: Zeeman Berry curvature. Quantum metrics: i) CQM: conventional quantum metric, ii) SRQM: spin-rotation quantum metric, and iii) ZQM: Zeeman quantum metric. ZMC: Zeeman metric connection, and ZSC: Zeeman symplectic connection.}
\label{tab:Symmetry_QGT}
\end{table}
%%%%%%%%%%---------------------------------------------------------------

\section{Quantum Kinetic theory of nonlinear Gyrotropic magnetotransport}
In this section, we derive the general expressions for the nonlinear gyrotropic magnetic (NGM) currents, which are governed by the generalized QGT, within the framework of quantum kinetic theory. In order to derive NGM currents, we start with the quantum Liouville equation ~\cite{culcer_2017}, where the time evolution of the single-particle density matrix ($\rho$) is governed by:
%%%%%%%%%%-------------------------------------------------------------
\begin{equation}
\frac{d\rho}{dt} + \frac{i}{\hbar}[\mathcal{H},\rho] = 0,
\end{equation}
%%%%%%%%%%---------------------------------------------------------------
where the total Hamiltonian of the system is expressed as $\mathcal{H} = \mathcal{H}_0 + \mathcal{H}_I + \mathcal{H}_U$ where $\mathcal{H}_0$ denotes the unperturbed band Hamiltonian. Here, $\mathcal{H}_I= - g\mu_B \mathbf{B}(t) \cdot \boldsymbol{\sigma}$, represents Zeeman coupling due to external time-dependent magnetic field $\mathbf{B}(t)$ where $\mu_B$ is the Bohr magneton, $g$ is the Landé $g$-factor, $\boldsymbol{\sigma} = (\sigma_x, \sigma_y, \sigma_z)$ is the vector of the Pauli matrices and $\mathbf{B}(t) =\mathbf{B}_0 \cos(\omega t)$ with frequency $\omega$. $\mathcal{H}_U$ represents the disorder term of the Hamiltonian. In the weak disorder limit, and within the relaxation time approximation, the above equation can be rewritten as:
%%%%%%%%%%---------------------------------------------------------------
\begin{equation}\label{eq:DME}
\frac{d\rho}{dt} 
+ \frac{i}{\hbar} \left[\left(\mathcal{H}_0 + \mathcal{H}_I\right), \rho\right] 
+ \frac{1}{\tau} \left[\rho - \rho^{(0)}\right] = 0,
\end{equation}
%%%%%%%%%%---------------------------------------------------------------
where $\tau$ is the relaxation time and we ignore the momentum dependence of it for simplicity. We solve the kinetic equation perturbatively by expanding the single-particle density matrix in powers of the magnetic field: $\rho(\mathbf{k}, t) = \rho^{(0)} + \rho^{(1)}(\mathbf{k}, t) + \rho^{(2)}(\mathbf{k}, t) + \cdots$,
where $\rho^{(n)}(\mathbf{k}, t) \propto B^n$ and $\rho^{(0)}$ is the equilibrium single-particle density matrix. 

Now, the $n$-th order density matrix in the band basis can be obtained by the solving the following equation:
%%%%%%%%%%---------------------------------------------------------------
\begin{equation}
\frac{d\rho^{(n)}_{mp}}{dt} + \frac{i}{\hbar} \left[\mathcal{H}_0, \rho^{(n)}\right]_{mp} + \frac{\rho^{(n)}_{mp}}{\tau} = \frac{i g\mu_B}{\hbar} \mathbf{B}(t) \cdot \left[\boldsymbol{\sigma}, \rho^{(n-1)}\right]_{mp}\label{eq:nth_order_DME}
\end{equation}
%%%%%%%%%%---------------------------------------------------------------
where $m,p$ are band indices. By solving the above equation iteratively, the first-order and second-order density matrix can be calculated as:

%%%%%%%%%%---------------------------------------------------------------
\begin{align}
\rho^{(1)}_{mp} &= \frac{g \mu_B}{2} 
\sum_{\omega' = \pm\omega} 
f_{mp} \, \phi_{mp} (\omega') \, \sigma^c_{mp} B^c_0 \, e^{-i\omega' t}, \label{eq:rho1}\\[6pt]
\rho^{(2)}_{mp} &= - 
\frac{g \mu_B}{2}
\sum_{\omega' = \pm\omega,l} 
\phi_{mp} (2\omega') 
\left[ \sigma_{ml}^b \rho_{lp}^{(1)} - \rho_{ml}^{(1)} \sigma_{lp}^b \right]
B_0^b \, e^{-i\omega' t}, \label{eq:rho2}
\end{align}
%%%%%%%%%%---------------------------------------------------------------
where $b$ and $c$ are the spatial indices, $\rho^{(0)}_{mp} = \delta_{mp} f_m^0$, $f_{mp} = f_m^0 - f_p^0$, $\epsilon_{mp} = \epsilon_m - \epsilon_p$, $\phi_{mp} (\omega') = \frac{1}{\omega' - \epsilon_{mp} + i/\tau}$, and $f^0$ is the equilibrium Fermi-Dirac distribution function. As in this work, we are interested in the second-order gyrotropic magnetic response, we focus on the second-order density matrix only from now on. The second-order density matrix can be decomposed into diagonal and off-diagonal components: $\rho^{(2)}_{mp} = \rho^{(2)}_{mm} + \rho^{(2)}_{m \neq p}$, where

%%%%%%%%%%---------------------------------------------------------------
\begin{align}
\rho^{(2)}_{mm} &= -\left( \frac{g\mu_B}{2} \right)^2 \sum_{\substack{\omega' = \pm\omega,l}} \phi_{2\omega'}^{mm} \left[ \phi_{lm}({\omega'}) \sigma_{ml}^b \sigma_{lm}^c f_{lm} \right. \nonumber \\
&\quad - \left. \phi_{ml}({\omega'}) \sigma_{ml}^c \sigma_{lm}^b f_{ml} \right] B_0^b B_0^c e^{-i2\omega' t}, \nonumber\\
\rho^{(2)}_{mp} &= \left( \frac{g\mu_B}{2} \right)^2 \Bigg[ 
\sum_{\substack{\omega' = \pm\omega}} 
\phi_{mp} (\omega') \left( \sigma_{mm}^b - \sigma_{pp}^b \right) \sigma_{mp}^c f_{mp} \nonumber \\
&\quad + \sum_{\substack{\omega' = \pm\omega \\ m \neq p \neq l}} 
\left(\phi_{lp}({\omega'}) f_{lp} \sigma_{ml}^b \sigma_{lp}^c 
- \phi_{ml}({\omega'}) f_{ml} \sigma_{ml}^c \sigma_{lp}^b \right) \Bigg]  \nonumber \\
&\quad \times \phi_{mp} (2\omega') B_0^c B_0^b e^{-i2\omega' t}.
\end{align}
%%%%%%%%%%---------------------------------------------------------------
Interestingly, both $\rho^{(2)}_{mm}$ and $\rho^{(2)}_{mp}$ contain the SRQGT, which is absent in the linear-order~\cite{Xiang_2025_PRL,Chakraborti2025}. The second term of the off-diagonal density matrix captures the effects arising from processes involving more than two energy bands. For simplicity and concreteness, we ignore this term by restricting our analysis to two-band systems. 

The second-order gyrotropic magnetic current density is defined as $j_{a}^{(2)} = \sum_{\substack{pm}} \tilde{v}_{pm}^{a} \rho_{mp}^{(2)}$, where $a$ represents the direction of the current. In the eigenbasis of $\mathcal{H}_0$, the generalized velocity operator takes the form:

%%%%%%%%%%---------------------------------------------------------------
\begin{align}
\tilde{v}^a_{pm}(\mathbf{k}) 
= \hbar^{-1} \left( \delta_{pm} v_{pm}^{a} + i r^a_{pm} \, \epsilon_{pm} \right),  
\label{eq:velocity}
\end{align}
%%%%%%%%%%---------------------------------------------------------------
where the first (intraband) term in Eq.~(\ref{eq:velocity}) corresponds to the usual band velocity,  
while the second (interband) term involves the Berry connection,  
$r_{mp} = \langle u_{m\mathbf{k}}^\zeta | i \partial_k | u_{p\mathbf{k}}^\zeta \rangle$~\cite{Wilczek_1984,Qian_2025,Niu_2008}.
% The current can be systematically expanded as a power series in the magnetic field strength, $\mathbf{j} = \sum_n \mathbf{j}^{(n)}$, where $\mathbf{j}^{(n)} \propto B^n$. The second-harmonic (SH) component corresponds to the $2\omega$ Fourier mode of the current, denoted as $\mathbf{j}^{(2)}$.  
Now the second-order gyrotropic magnetic current density can be decomposed into diagonal and off-diagonal contributions, $\mathbf{j}^{(2)} = \mathbf{j}^{(d)} + \mathbf{j}^{(od)}$, where $d, od$ represent the diagonal and off-diagonal components respectively. The corresponding NGM conductivity is defined as $\chi_{abc}^{(2)} = j^{(2)}_a / (B^b B^c)$, which also naturally separates into $\chi_{abc} = \chi_{abc}^{(d)} + \chi_{abc}^{(od)}$.
%with $\chi_{abc}^{(d)}$ arising from the diagonal terms and $\chi_{abc}^{(od)}$ from the off-diagonal terms.  

Explicitly, the diagonal component of the NGM conductivity can be written as:
%%%%%%%%%%-------------------------------------------------------------
\bea
    \chi_{abc}^{(d)} &=& \left( \tfrac{g \mu_B}{2} \right)^2 \sum_{\substack{\omega' = \pm\omega, \\ mp}} \phi_{mm}(2\omega') \, v_{mp}^a \non\\
    &\times&\left( \mathscr{R}_{mp}^{bc} - \tfrac{i}{2} \Lambda^{bc}_{mp} \right) \, \phi_{mp}(\omega') f_{pm}, 
\eea
%%%%%%%%%%-------------------------------------------------------------
where $v_{mp}^a = v_{mm}^a - v_{pp}^a$. It is clear that the diagonal component is entirely governed by the SRQGT, which incorporates both the spin rotation Berry curvature and the spin rotation quantum metric. Now the off-diagonal component of NGM conductivity can be defined as:
%%%%%%%%%%-------------------------------------------------------------
\bea
    \chi_{abc}^{(od)} &=& i \left( \tfrac{g \mu_B}{2} \right)^2 \sum_{\substack{\omega' = \pm\omega, \\ mp}} \phi_{mp}({2\omega'}) \, \epsilon_{mp} \non\\
    &\times&\left( \mathscr{L}_{mp}^{abc} - i \tilde{\mathscr{L}}_{mp}^{abc} \right) {\phi}_{mp}({\omega'}) f_{pm}.  
\eea
%%%%%%%%%%-------------------------------------------------------------
The off-diagonal component is goverened by the two novel geometric quantities: the Zeeman metric connection $\mathscr{L}_{mp}^{abc}$ and the Zeeman symplectic connection $\tilde{\mathscr{L}}_{mp}^{abc}$. These quantities arise through the relation  
$r_{pm}^a \, \mathcal{D}_{mp}^b \, \sigma_{mp}^c = \mathscr{L}_{mp}^{abc} - i \, \tilde{\mathscr{L}}_{mp}^{abc}$, where $\mathcal{D}_{mp}^b = \sigma_{mm}^b - \sigma_{pp}^b$. Taking symmetry into account, the Zeeman symplectic connection $\tilde{\mathscr{L}}_{mp}^{abc}$ is odd under both $\hat{\mathcal{P}}$ and $\hat{\mathcal{T}}$, whereas the Zeeman metric connection $\mathscr{L}_{mp}^{abc}$ is odd under $\hat{\mathcal{P}}$ but even under $\hat{\mathcal{T}}$.

Further, the second-order gyrotropic magnetoconductivity can be written as the sum of conduction and displacement conductivity components, $\chi^{(2)}_{abc} = \chi_{abc}^C + \chi_{abc}^D$. In the low-frequency regime well below the interband absorption threshold, i.e.\ $\hbar \omega \ll \epsilon_{pm}$, and in the clean limit $(\tau \to \infty)$, the quantity $\chi_{abc}^{(2)}$ can be further simplified as:

%%%%%%%%%%-------------------------------------------------------------
\begin{align}
    \chi_{abc}^{C,od} &= \left(\frac{g\mu_B}{2}\right)^2 
    \sum_{mp} f_{m} \left(\frac{4}{\epsilon_{mp}}\right) 
    \tilde{\mathscr{L}}_{mp}^{abc}, \notag \\
    \chi_{abc}^{D,od} &= \left(\frac{g\mu_B}{2}\right)^2 
    \sum_{mp} f_{m} \left(\frac{12\hbar\omega}{\epsilon_{mp}^2}\right) 
    \mathscr{L}_{mp}^{abc}, \notag \\
    \chi_{abc}^{C,d} &= \left(\frac{g\mu_B}{2}\right)^2 
    \sum_{mp} f_{m}\,\frac{\partial \epsilon_{mp}}{\partial k^a} 
    \left(\frac{\mathscr{R}_{mp}^{bc}}{\epsilon_{mp}^2}\right), \notag \\
    \chi_{abc}^{D,d} &= \left(\frac{g\mu_B}{2}\right)^2 
    \sum_{mp} f_{m}\,\frac{\partial \epsilon_{mp}}{\partial k^a} 
    \left(\frac{\Lambda^{ab}_{mp}}{\hbar\omega \epsilon_{mp}}\right). 
    \label{eq:conductivities}
\end{align}
%%%%%%%%%%-------------------------------------------------------------
% A detailed derivation of these expressions is provided in the \ttd{Appendix}.
Since the conduction and displacement NGM conductivities derived here are independent of the scattering time $\tau$, we define them as intrinsic nonlinear gyrotropic magneto-conductivities. Importantly, this intrinsic response is determined solely by the properties of the Bloch wavefunctions, rendering it robustness even in the presence of weak disorder. $\chi_{abc}^{C,\mathrm{od}}$ is formally analogous to the shift-conductivity, while $\chi_{abc}^{D,\mathrm{d}}$ is analogous to the injection-conductivity~\cite{kamal_2022_prl}. The tensors $\chi_{abc}^{C,od}$ and $\chi_{abc}^{D,d}$ are antisymmetric with respect to the indices $b$ and $c$, owing to the antisymmetric nature of $\tilde{\mathscr{L}}_{mp}^{abc}$ and $S_{mp}^{bc}$ respectively. In contrast, $\chi_{abc}^{D,od}$ and $\chi_{abc}^{C,d}$ are symmetric under the exchange of $b$ and $c$, as $\mathscr{L}_{mp}^{abc}$ and $\mathscr{R}_{mp}^{bc}$ itself are symmetric in these indices.

%In comparison with Ref.~\cite{kamal_2022_prl}, similar types of currents appear at second order, although their work considers an electric-field expansion whereas our results arise at second order in the magnetic field. The diagonal conductivity exhibits identical symmetry properties in both cases, as summarized in Table~\ref{tab:Symmetry_table}, since the SQGT and CQGT transform in the same way under $\hat{\mathcal{T}}$ and $\hat{\mathcal{P}}$. The key distinction is that, for a $\hat{\mathcal{T}}$-symmetric system, their analysis leads to a surviving symplectic connection, whereas in our case it is the Zeeman quantum metric connection that remains since the ZQGT and CQGT transform in the different way under $\hat{\mathcal{T}}$ and $\hat{\mathcal{P}}$ .

\section{Results}

In this section, we establish a systematic framework to demonstrate how symmetry breaking and band geometry generate NGM currents. Specifically, we consider four representative systems: (i) a massless Dirac system preserving time-reversal ($\mathcal{T}$) but breaks $\mathcal{P}$ and combined $\mathcal{PT}$; (ii) a Dirac system with hexagonal warping, which also preserves $\mathcal{T}$ but breaks $\mathcal{P}$ and combined $\mathcal{PT}$, (iii) a tilted massive Dirac system in which $\mathcal{P}$, $\mathcal{T}$, and $\mathcal{PT}$ symmetries are broken; and (iv) the antiferromagnet CuMnAs, where $\mathcal{P}$ and $\mathcal{T}$ are individually broken while the combined $\mathcal{PT}$ symmetry remains intact.

\subsection{Two Dimensional Dirac System}

To study the behavior of NGM conductivity, we first consider a generic model of a two-dimensional (2D) Dirac system, whose effective Hamiltonian captures the essential physics of materials hosting Dirac fermions. The band structure of the system exhibits a isotropic Dirac cone, where the conduction and valence bands intersect linearly at a single $\mathbf{k}$-point, giving rise to gapless excitations with linear dispersion. The corresponding low-energy Hamiltonian is given by:
%%%%%%%%%%-------------------------------------------------------------
\begin{align}
H(\mathbf{k}) = v_F (k_x \sigma_y -  k_y \sigma_x),
\end{align}
%%%%%%%%%%-------------------------------------------------------------
where $v_F$ is the Fermi velocity, and $\boldsymbol{\sigma} = (\sigma_x, \sigma_y, \sigma_z)$ are the Pauli matrices acting on the spin degree of freedom. The band dispersion is $\epsilon^{\pm}_{\mathbf{k}} = \pm v_F k$ with $k=\sqrt{k_x^2 + k_y^2}$, corresponding to gapless Dirac cone. In the context of symmetry this Hamiltonian preserves TRS symmetry but explicitly breaks inversion symmetry, since inversion reverses momentum while leaving spin unchanged. It remains invariant under the mirror operation $\mathcal{M}_y: (y\rightarrow -y)$ and $\mathcal{M}_x: (x\rightarrow -x)$, represented by $\mathcal{M}_y = i\sigma_y$ and $\mathcal{M}_x = i\sigma_x$ respectively. Furthermore, the isotropy of the Fermi velocity ensures that the system retains the full $C{\infty v}$ point-group symmetry.

Table~\ref{tab:NGMC_summary} presents the analytical expressions for the various components of the QGT that govern the NGM conductivity.
% %%%%%%%%%%-------------------------------------------------------------
% \begin{align}
% &S_{\pm \mp}^{xy} = 0 = S_{\mp \pm}^{yx} = S_{\mp \pm}^{xx} = S_{\mp \pm}^{yy}, \nonumber \\[6pt]
% &T_{\pm \mp}^{xx/yy} = \mp \frac{v_F^2 k_{x/y}^2}{R^2}, \quad
% T_{\pm \mp}^{xy} = \pm \frac{v_F^2 k_x k_y}{k^2} = T_{\pm \mp}^{yx}, \nonumber \\[6pt]
% &\tilde{L}_{\pm \mp}^{xxx} = \mp\frac{v_F^2 k_x k_y^2}{2k^2} = \mp \tilde{L}_{\pm \mp}^{xyy}, \quad
% \tilde{L}_{\pm \mp}^{xxy} = \mp\frac{v_F^2 k_y^3}{2k^2}, \nonumber \\[6pt]
% &L_{\pm \mp}^{xxy} = 0 = L_{\pm \mp}^{xxx} = L_{\pm \mp}^{xyx}.
% \label{eq:QGT_2D}
% \end{align}
% %%%%%%%%%%-------------------------------------------------------------
The components $\mathscr{R}_{\pm \mp}^{xx}$ and $\mathscr{R}_{\pm \mp}^{yy}$ scale as $k_x^2$ and $k_y^2$ respectively and remain sign-definite near the $\Gamma$ point, reflecting their monopolar character. In contrast, $\mathscr{R}_{\pm \mp}^{xy}$, which scales as $k_x k_y$, exhibits a quadrupolar nature. On the other hand, $\tilde{\mathscr{L}}$ is an odd function of either $k_x$ or $k_y$. Consequently, both $\mathscr{L}_{\pm \mp}$ and $\Lambda_{\pm \mp}$ vanish. 

From the above expressions of the QGT, it is evident that the displacement NGM conductivity vanishes, as it is governed by the Zeeman metric connection ($\mathscr{L}$) and the spin quantum metric ($\mathscr{R}$). The integrand of the conduction NGM conductivity is an odd function of $k_x$ and $k_y$, leading to its cancellation upon integration over momentum space. This behavior can be understood from the presence of TRS, which imposes strict constraints on the conduction part of the NGM conductivity, as summarized in Table~\ref{tab:Symmetry_QGT}. Therefore, for this TRS-symmetric 2D Dirac model, the NGM conductivity vanishes identically.

To generate finite NGM currents, we consider three distinct symmetry-broken realizations of Dirac systems: (i) surface states of 3D topological insulators (TIs) with hexagonal warping, (ii) tilted massive Dirac systems and (iii) the antiferromagnet CuMnAs.

\subsection{Surface states of TI with hexagonal warping}

Building upon the discussion of the 2D massless Dirac system, we now investigate the influence of crystalline anisotropy on the NGM conductivity by incorporating hexagonal warping. To this end, we consider a 2D Dirac system subject to a hexagonal warping term, described by the low-energy Hamiltonian~\cite{Fu2009,Hasan2010,Liang_2009}:
%%%%%%%%%%---------------------------------------------------------------
\begin{align}
H^{\rm HW} &= H(\mathbf{k}) + \frac{\lambda}{2}\left(k_+^3 + k_-^3\right)\sigma_z,
\label{eq:warping_ham}
\end{align}
%%%%%%%%%%---------------------------------------------------------------
where $k_\pm = k_x \pm i k_y$, and $\lambda$ denotes the strength of the hexagonal warping. The hexagonal warping term describes cubic spin-orbit coupling at the surface of rhombohedral crystal systems. It preserves TRS but breaks inversion symmetry. It also breaks continuous rotational invariance, reducing it to the discrete threefold rotational symmetry $C_{3z}$ about the $z$ axis~\cite{Fu2009}, and explicitly violates the mirror symmetry $\mathcal{M}_y$, while preserving $\mathcal{M}_x$. The corresponding energy dispersion is given by $\epsilon^{\pm}_{\mathbf{k}} = \pm \sqrt{v_F^2 k^2 + \lambda^2 k^6 \cos 3\theta}$
where $\theta = \tan^{-1}(k_y/k_x)$, and the $+$ ($-$) sign corresponds to the conduction (valence) band. The dispersion exhibits a sixfold rotational symmetry under $\theta \rightarrow \theta + \pi/3$. In the absence of warping ($\lambda = 0$), the Fermi surface is perfectly circular. Upon introducing the warping term, the Fermi surface remains nearly circular for small $\lambda$, but as the warping strength increases, it gradually deforms into a noncircular, snowflake-like shape—developing sharp protrusions along high-symmetry directions and concave regions in between.

% The components of QGT can be evaluated analytically as:
% %%%%%%%%%%---------------------------------------------------------------
% \begin{align}
% S_{\pm \mp}^{xy} &= \mp\frac{2\lambda\left(k_x^3 - 3k_x k_y^2\right)}{R}, \quad  
% T_{\pm \mp}^{xy} =\pm \frac{v_F^2 k_x k_y}{R^2},\nonumber\\[2pt]
% T_{\pm \mp}^{xx} &= \pm\frac{4\lambda^2\left(k_x^3 - 3k_x k_y^2\right)^2 + k_x^2 v_F^2}{R^2}, 
% \quad S_{\pm \mp}^{xx} = 0,\nonumber \\[2pt]
% L_{\pm \mp}^{xxx} &=\pm \frac{v_F^2 \lambda k_y k_x^3}{R^3}, \quad
% L_{\pm \mp}^{xxy} =\mp \frac{3 \lambda v_F^2 k_y^2 (k_x^2 - k_y^2)}{2 R^3}, \nonumber\\[2pt]
% \tilde{L}_{\pm \mp}^{xxy} &=\mp \frac{v_F^2 k_y \left[3\lambda^2k_x^2 (k_x^2 - 3k_y^2) \delta 
% + \lambda^3 k_x^2 (k_x^2 - 3k_y^2)^2 \right]}{2 R^4},\nonumber \\[2pt] 
% \tilde{L}_{\pm \mp}^{xxx} &=\mp \frac{v_F^2 k_x k_y^2\left(v_F^2 + 3\lambda^2 (k_x^2-3k_y^2)\delta\right)}{2 R^4},\nonumber  
% \end{align}
% %%%%%%%%%---------------------------------------------------------------
% with $R = \sqrt{v_F^2 k^2 + \frac{\lambda^2}{4}\left(k_+^3 + k_-^3\right)^2}, \quad \delta = k_x^2 - k_y^2$.

All the analytical expression of QGT is given in~\ref{tab:NGMC_summary}. The off-diagonal component of spin quantum metric $\mathscr{R}^{xy}_{mp} \propto k_x k_y$ changes sign under a $90^\circ$ rotation in momentum space, indicating a quadrupolar-like structure. 
In contrast, the diagonal component $\mathscr{R}_{\pm \mp}^{xx} \propto k_x^2$ remains sign-definite under such rotations, exhibiting a monopolar-like structure. 
Moreover, $\Lambda_{\pm \mp}^{xy}$, being odd in $k_x$, manifests a dipolar character, highlighting its directional asymmetry. 
Furthermore, the components $\mathscr{L}_{\pm \mp}^{xxx}$ and $\mathscr{L}_{\pm \mp}^{xxy}$ are even functions of $(k_x, k_y)$, consistent with a monopole-like structure that is enforced by time-reversal symmetry whereas the other components $\tilde{\mathscr{L}}_{\pm \mp}$ are odd in $(k_x, k_y)$, reflecting dipolar-like structures.

%%%%%%%%%%---------------------------------------------------------------
\begin{figure}[ht!]
\centering
\includegraphics[width=\columnwidth]{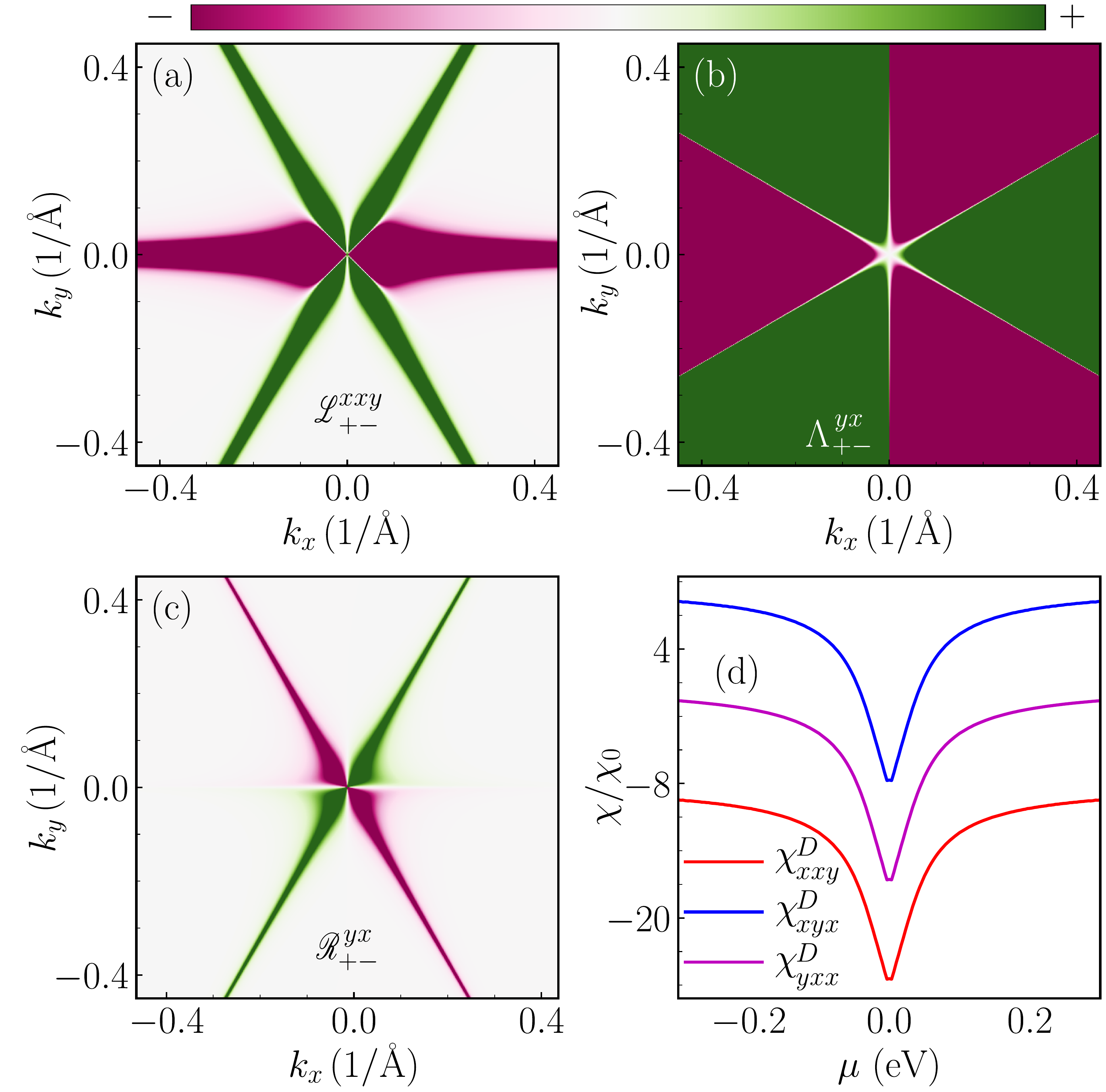}
\caption{(a) The $xxy$ component of the Zeeman metric connection is plotted, exhibiting a nearly monopolar character that gives rise to a NGM conductivity.
(b) The spin-rotation Berry curvature is shown, displaying a dipolar structure that also contributes to a nonvanishing NGM conductivity.
(c) The spin-rotation quantum metric is plotted, revealing a quadrupolar character that results in a vanishing NGM conductivity. This behavior is consistent with time-reversal symmetry (TRS), under which the conduction NGM conductivity must vanish.
(d) The chemical potential ($\mu$) dependence of the nonvanishing independent components of the displacement NGM conductivity in a hexagonally warped Dirac system is shown. The parameters used are $v_f = 1,\text{eV}$, $\lambda = 255,\text{eV}\cdot\text{\AA}^3$, $T = 10 \text{K}$ and $\chi_0 = \left(\frac{g\mu_B}{2}\right)^2 \,\mathrm{A\,m^{-1}\,T^{-2}}$.}
\label{warping}
\end{figure}
%%%%%%%%%%--------------------------------------------------------------- 

Using the different components of the QGT, we evaluate the NGM conductivity and find that the contribution of conduction NGM conductivity vanishes as a consequence of TRS preservation (see Table~\ref{tab:Symmetry_QGT}). In contrast, the displacement currents remain finite, consistent with TRS symmetry (Table~\ref{tab:Symmetry_QGT}). The presence of $\mathcal{M}_x$ mirror symmetry enforces that all components of the displacement NGM conductivity containing an odd number of $x$ indices vanish. We identify three independent components of the displacement NGM conductivity tensor, $\chi_{xxy}$, $\chi_{xyx}$, and $\chi_{yxx}$ with the relation $\chi_{yyy} = -\chi_{yxx}$. Interestingly, the diagonal and off-diagonal components of $\chi_{xxy}$ and $\chi_{xyx}$ satisfy $\chi_{xyx}^{D,d} = -\chi_{xxy}^{D,d}$ and $\chi_{xyx}^{D,od} = \chi_{xxy}^{D,od}$. This behavior can be understood from the symmetry properties of the underlying QGTs. The diagonal component $\chi^{D,d}$ is governed by $\Lambda_{mp}^{bc}$, which is antisymmetric under the exchange of indices $b$ and $c$, i.e., $\Lambda_{mp}^{xy} = -\Lambda_{mp}^{yx}$, ensuring that $\chi_{xyx}^{D,d} = -\chi_{xxy}^{D,d}$. In contrast, the off-diagonal component $\chi^{D,od}$ is governed by $\mathscr{L}_{mp}^{abc}$, which is symmetric under the exchange of $b$ and $c$, ensuring that $\chi_{xyx}^{D,od} = \chi_{xxy}^{D,od}$. The independent components of the NGM conductivity tensor are plotted in Fig.~\ref{warping}. All components exhibit a pronounced peak at the Dirac point ($\mu = 0$), which originates from the presence of $\epsilon_{mp}$ in the denominator of Eq.~\ref{eq:conductivities}. Near the Dirac point, $\epsilon_{mp}$ becomes vanishingly small, resulting in a strong enhancement of the response. As the chemical potential is tuned away from $\mu = 0$, the components decrease symmetrically on either side of the Dirac point. The smooth, continuous variation of these components further reflects the persistent TRS of the system.

We would like to emphasize that the [111] surface states of Bi$_2$Te$_3$~\cite{Fu2009} are accurately described by the Hamiltonian given in Eq.~(\ref{eq:warping_ham}). Hence, this material provides a realistic platform for experimentally verifying the proposed results.

\begingroup
\setlength{\tabcolsep}{6.2pt}
\begin{table*}[!ht]
\centering
\caption{The summary of systems, their symmetries, corresponding quantum geometric tensor (QGT) characteristics, 
and nonlinear gyrotropic magnetic (NGM) conductivity components is presented in the table below. 
Here, we define $l_k = \sqrt{v_F^2 k^2 + \frac{\lambda^2}{4} (k_+^3 + k_-^3)^2}$, 
$\delta = (k_x^2 - k_y^2)$, and $m_k = \sqrt{v_F^2 k^2 + \Delta^2}$. $U$ and $V$ are some complicated functions of $\mathbf{k}$.}
\label{tab:NGMC_summary}
\begin{tabular}{c c c c}
\hline
\textbf{System} & \textbf{Symmetry} & \textbf{Quantum Geometry} & \textbf{\shortstack{Independent NGM \\ conductivities }} \\
\hline
2D massless Dirac & \shortstack{$\hat{\mathcal{T}}$, $\mathcal{M}_x$,\\ $\mathcal{M}_y$, $C_{\infty v}$} &
\(
\begin{aligned}
&\Lambda_{\pm \mp}^{xy} = \Lambda_{\mp \pm}^{yx} = \Lambda_{\mp \pm}^{xx} = \Lambda_{\mp \pm}^{yy} = 0,\\
&\mathscr{R}_{\pm \mp}^{xx/yy} = \mp \frac{v_F^2 k_{x/y}^2}{k^2}, \quad
\mathscr{R}_{\pm \mp}^{xy/yx} = \pm \frac{v_F^2 k_x k_y}{k^2},\\
&\tilde{\mathscr{L}}_{\pm \mp}^{xxx} = \mp \frac{v_F^2 k_x k_y^2}{2 k^2}, \quad
\tilde{\mathscr{L}}_{\pm \mp}^{xyy} = \mp \frac{v_F^2 k_x k_y^2}{2 k^2}, \\
&\tilde{\mathscr{L}}_{\pm \mp}^{xxy} = \mp \frac{v_F^2 k_y^3}{2 k^2},\\
&\mathscr{L}_{\pm \mp}^{xxy} = \mathscr{L}_{\pm \mp}^{xxx} = \mathscr{L}_{\pm \mp}^{xyx} = 0
\end{aligned}
\) & $\times$ \\
\hline
$\shortstack{\text{Surface states of TI} \\ \text{with hexagonal warping}}$ &  $\hat{\mathcal{T}}$, $\mathcal{M}_x$, $C_3$ & 
\(
\begin{aligned}
&\Lambda_{\pm \mp}^{xy} = \mp \frac{2\lambda\left(k_x^3 - 3k_x k_y^2\right)}{l_k}, \quad  
\mathscr{R}_{\pm \mp}^{xy} =\pm \frac{v_F^2 k_x k_y}{l_k^2},\\[2pt]
&\mathscr{R}_{\pm \mp}^{xx} = \pm\frac{4\lambda^2\left(k_x^3 - 3k_x k_y^2\right)^2 + k_x^2 v_F^2}{l_k^2}, 
\quad \Lambda_{\pm \mp}^{xx} = 0,\\[2pt]
&\mathscr{L}_{\pm \mp}^{xxx} =\pm \frac{v_F^2 \lambda k_y k_x^3}{l_k^3}, \quad
\mathscr{L}_{\pm \mp}^{xxy} =\mp \frac{3 \lambda v_F^2 k_y^2 (k_x^2 - k_y^2)}{2 l_k^3},\\[2pt]
&\tilde{\mathscr{L}}_{\pm \mp}^{xxy} =\mp \frac{v_F^2 k_y \left[3\lambda^2k_x^2 (k_x^2 - 3k_y^2) \delta 
+ \lambda^3 k_x^2 (k_x^2 - 3k_y^2)^2 \right]}{2 l_k^4},\\[2pt] 
&\tilde{\mathscr{L}}_{\pm \mp}^{xxx} =\mp \frac{v_F^2 k_x k_y^2\left(v_F^2 + 3\lambda^2 (k_x^2-3k_y^2)\delta\right)}{2 l_k^4}
\end{aligned}
\) & $\chi_{xxy}^D$, $\chi_{xyx}^D$, $\chi_{yxx}^D$ \\

\hline
Tilted massive Dirac & $\times$ &
\(
\begin{aligned}
&\Lambda_{\pm \mp}^{xy} =\mp \frac{\Delta}{m_k} = \Lambda_{\mp \pm}^{yx}, \quad  
\mathscr{R}_{\pm \mp}^{xy} =\pm \frac{v_F^2 k_x k_y}{m_k^2} = \mathscr{R}_{\pm \mp}^{yx},\\[6pt]
&\mathscr{R}_{\pm \mp}^{xx} =\mp \frac{v_F^2 k_x^2 + \Delta^2}{m_k^2} = \mathscr{R}_{\pm \mp}^{yy}, \quad
S_{\pm \mp}^{xx} = 0,\\[6pt]
&\mathscr{L}_{\pm \mp}^{xxx} = \mp\frac{v_F^2 k_y \Delta}{2 m_k^3}, \quad
\tilde{\mathscr{L}}_{\pm \mp}^{xxx} = \mp\frac{v_F^2 k_x k_y^2}{2m_k^4},\\[6pt]
&\mathscr{L}_{\pm \mp}^{xxy} = 0, \quad
\tilde{\mathscr{L}}_{\pm \mp}^{xxy} = \mp\frac{v_F^2 k_y \left(\beta^2 + v_F^2 k_y^2\right)}{2m_k^4}
\end{aligned}
\) & \shortstack{$\chi_{xxy}^C$, $\chi_{yxx}^C$, $\chi_{xyx}^C$,\\ $\chi_{xxx}^D$, $\chi_{yxy}^D$} \\
\hline
$\mathcal{PT}$-symmetric CuMnAs & $\mathcal{PT}$ & 
\(
\begin{aligned}
&\tilde{\mathscr{L}}_{\pm \mp}^{abc} = U(k_x,k_y), \mathscr{R}_{\pm \mp}^{ab} = V(k_x,k_y),\Lambda_{\pm\mp}^{ab} = \mathscr{L}_{\pm \mp}^{abc} = 0
\end{aligned}
\)& \shortstack{$\chi_{xxy}^C$, $\chi_{xyx}^C$,\\ $\chi_{yyy}^C$, $\chi_{yxx}^C$} \\
\hline
\end{tabular}
\end{table*}
\endgroup

\subsection{Tilted Massive Dirac System}

We consider a generic model of a tilted 2D massive Dirac cone described by
%%%%%%%%%%-------------------------------------------------------------
\begin{align}
H^{\mathrm{TMD}}(\mathbf{k})
= H(\mathbf{k}) + t k_y + \Delta \sigma_z ,
\end{align}
%%%%%%%%%%-------------------------------------------------------------
where $t$ denotes the tilt parameter along the $k_y$ direction, and $\Delta$ represents the mass term that opens a gap in the system. The corresponding energy spectrum is given by $\epsilon^\pm_{\mathbf{k}} = t k_y \pm \sqrt{v_F^2 k^2 + \Delta^2}$ where the $+$ ($-$) sign refers to the conduction (valence) band. The tilt term deforms the Dirac cone and breaks particle–hole symmetry, while also explicitly violating TRS.

Under the mirror operations $\mathcal{M}_x$ and $\mathcal{M}_y$, the tilt term $t k_y$ is odd under $\mathcal{M}_y$, whereas the mass term $\Delta \sigma_z$ is odd under both $\mathcal{M}_x$ and $\mathcal{M}_y$. Consequently, both mirror symmetries are broken. The isotropic Dirac term is odd under inversion, implying that inversion symmetry is already absent and remains broken upon inclusion of the tilt and mass terms. Furthermore, the tilt term $t k_y$ breaks the rotational symmetry that characterizes the original 2D Dirac Hamiltonian, both with and without warping. Experimentally, tilted Dirac cones have been observed in organic conductors such as $\alpha$-(BEDT-TTF)$_2$I$_3$~\cite{Goerbig_2008}, and have also been theoretically predicted in transition-metal dichalcogenides such as WTe$_2$~\cite{Li_2017}.

The QGTs governing the NGM conductivities have been calculated analytically and given in Table~\ref{tab:NGMC_summary}.
%%%%%%%%%%---------------------------------------------------------------
\begin{figure}[ht!]
\centering
\includegraphics[width=\columnwidth]{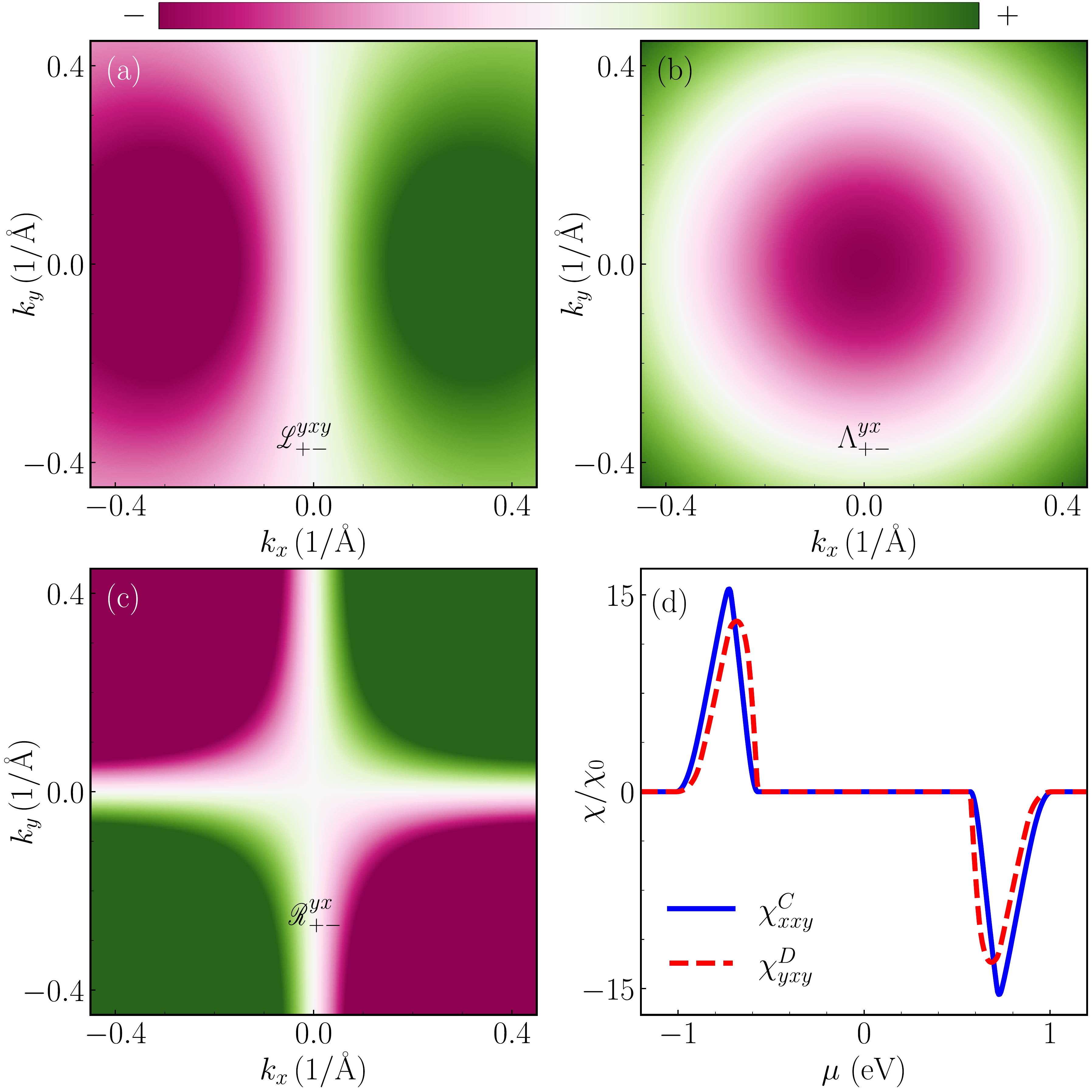}
\caption{(a) The $yxy$ component of the Zeeman metric connection is plotted, exhibiting a  dipolar behaviour.
(b) The spin-rotation Berry curvature displays a nearly monopolar structure and
(c) the off-diagonal component of spin-rotation quantum metric shows a quadrupolar character.
(d) The chemical potential ($\mu$) dependence of the nonvanishing components of the conduction and displacement NGM conductivity is presented for a tilted massive Dirac system. The parameters used are $v_f = 1~\text{eV}$, $\Delta = 0.6~\text{eV}$, $t = 0.3~\text{eV}$, and $\chi_0 = \left(\frac{g\mu_B}{2}\right)^2 \,\mathrm{A\,m^{-1}\,T^{-2}}$.}
\label{tilted}
\end{figure}
%%%%%%%%%%---------------------------------------------------------------
For instance, the off-diagonal component of SRQM $\mathscr{R}^{xy}_{mp} \propto k_x k_y$ changes sign under a $90^\circ$ rotation in $k$-space, indicating a quadrupolar nature. In contrast, the diagonal component of SRQM $\mathscr{R}^{xx}_{mp}$ and off-diagonal component of SBC $\Lambda^{xy}_{mp}$ scale as $k_x^2$ and remain sign-definite near the $\Gamma$ point, reflecting monopolar character. Furthermore $\mathscr{L}_{\pm \mp}^{xxx}$, $\tilde{\mathscr{L}}_{\pm \mp}^{xxx}$, and $\tilde{\mathscr{L}}_{\pm \mp}^{xxy}$ are odd in $(k_x,k_y)$, behaving like dipoles. The component $\mathscr{L}_{\pm \mp}^{xxy} = 0$ vanishes owing to the absence of the corresponding real part.

Using the various components of the QGT, we evaluate the NGM conductivity and find that both the conduction and displacement contributions vanish when the tilt parameter $t = 0$. A finite NGM conductivity emerges only when $t \neq 0$, indicating that it arises due to the breaking of particle–hole symmetry.
We identify three independent components of the conduction NGM conductivity tensor, $\chi_{xxy}$, $\chi_{yxx}$, and $\chi_{xyx}$ and two independent components of displacement NGM conductivity $\chi_{xxx}$ and $\chi_{yxy}$. Interestingly, the diagonal and off-diagonal components of different NGM conductivity satisfy $\chi_{xxy}^{C,d} = \chi_{xyx}^{C,d}$, $\chi_{yyy}^{C,d} = \chi_{yxx}^{C,d}$, $\chi_{xxy}^{C,od} = -\chi_{yxx}^{C,od}$, $\chi_{xyx}^{C,od} = -\chi_{yyy}^{C,od}$, $\chi_{yxy}^{D,d} = - \chi_{yyx}^{D,d}$ and $\chi_{yxy}^{D,od} = \chi_{xxx}^{D,od}$. This behavior can be understood from the symmetry properties of the underlying QGTs. The diagonal component $\chi^{D,d}$ is governed by SRBC $\Lambda^{ab}_{mp}$, which is antisymmetric under the exchange of indices $b$ and $c$, ensuring that $\chi_{yxy}^{D,d} = -\chi_{yyx}^{D,d}$.  In contrast the diagonal component $\chi^{C,d}$ is governed by SRQM $\mathscr{R}^{bc}_{mp}$, which is symmetric under the exchange of indices $b$ and $c$, ensuring that $\chi_{xxy}^{C,d} = \chi_{xyx}^{C,d}$. The equality $\chi_{yyy}^{C,d} = \chi_{yxx}^{C,d}$ arises from the isotropic nature of the tensor component of spin quantum metric $\mathscr{R}^{yy}_{mp} = \mathscr{R}^{xx}_{mp}$. Similarly, the relation $\chi_{yxy}^{D,od} = \chi_{xxx}^{D,od}$ originates from the symmetric property of the ZQM, $\mathcal{F}^{xx}_{mp} = \mathcal{F}^{yy}_{mp}$. In contrast, $\chi_{xxy}^{C,od} = -\chi_{yxx}^{C,od}$ stems from the antisymmetric property of the ZBC, which satisfies $\Gamma^{ab}_{mp} = -\Gamma^{ab}_{mp}$. $\chi_{xyx}^{C,od} = -\chi_{yyy}^{C,od}$ originates from the asymmetry of ZBC $\Gamma_{mp}^{xx} = - \Gamma_{mp}^{yy}$.
The NGM conductivity components $\chi_{xxy}^{C}$ and $\chi_{yxy}^{D}$ are depicted in Fig.~\ref{tilted}. These components vanish when the chemical potential lies within the band gap, indicating the absence of available charge carriers. As $\mu$ is tuned into the conduction or valence bands, they become finite and exhibit antisymmetric peaks, reflecting opposite contributions from electron and hole excitations on either side of the Dirac point.

% The analytical expression of some of the non vanishing 

% %%%%%%%%%%---------------------------------------------------------------
% \begin{align}
%     \sigma_{yxy}^{C,d} &= 
%     \frac{\pi t \sqrt{\mu^2 - \Delta^2}\,(\mu^2 + v_F^2)}{\mu^3}, \nonumber\\[6pt]
%     \sigma_{yxy}^{C,od} &= 
%     \frac{\pi t}{24} \, \frac{\sqrt{v_F(\mu^2 - \Delta^2)} \, (\Delta^2 + 8\mu^2)}{\mu^3}, \nonumber \\[6pt]
%     \sigma_{xxx}^{D,od} &= 
%     \frac{ \pi t}{24} \, \frac{(\mu^2 - \Delta^2)^{3/2}}{\mu^3},\nonumber
% \end{align}
% %%%%%%%%%%---------------------------------------------------------------
% All three components scale linearly with the tilt parameter $t$, underscoring that tilt is indispensable: in its absence, the NGMC vanishes identically. Hence, tilt not only reshapes the Dirac cone geometry but also serves as the fundamental origin of the nonlinear gyrotropic magnetoconductivities.

The above analysis highlights the geometric origin of nonlinear gyrotropic magnetic responses induced by band tilting. A related phenomenon emerges on the surface of topological crystalline insulators (TCIs). For instance, the [001] surface of SnTe and Pb$_{1-x}$Sn$_x$X (with $X =$ Se, Te) hosts four massless Dirac fermions protected by two mirror symmetries. At low temperatures, the surface undergoes a structural transition into a ferroelectric phase, during which one of the mirror symmetries is spontaneously broken while the other remains intact. Consequently, two of the surface Dirac cones acquire a finite mass gap, whereas the remaining two stay massless. Since the massless Dirac cones yield a vanishing NGM conductivity, as discussed earlier, it suffices to consider the contributions from the two massive Dirac cones in the distorted crystal phase. The corresponding low-energy Hamiltonian around the Dirac nodes can be expressed as~\cite{Sodemann_2015}:
%%%%%%%%%%-------------------------------------------------------------
\begin{align}
H^{\rm TCI}_{\eta} = v_x k_x \sigma_y - \eta v_y k_y \sigma_x + \eta t k_y + \Delta \sigma_z,
\end{align}
%%%%%%%%%%-------------------------------------------------------------
where $\eta = \pm 1$ denotes the valley index. Although inversion symmetry is broken, the two Dirac cones are related by time-reversal symmetry, ensuring that the total conduction NGM conductivity vanishes due to this effective TRS. Specifically, the conduction components $\chi_{abc}^{C, d}$ and $\chi_{abc}^{C, od}$ exhibit opposite signs under the transformation $\eta \rightarrow -\eta$, leading to complete cancellation when contributions from both valleys are summed. In contrast, the displacement component of $\chi_{abc}^{D, od}$ remains invariant under valley exchange and therefore survives in the net response.

\subsection{\texorpdfstring{$\mathcal{P}\mathcal{T}$}{PT}-symmetric CuMnAs}

We now turn to the $\mathcal{PT}$-symmetric antiferromagnet CuMnAs, which provides a realistic platform for intrinsic nonlinear gyrotropic magnetic responses governed by quantum geometry. CuMnAs crystallizes in a bipartite lattice structure with antiferromagnetic ordering, where opposite spins reside on different sublattices. While both $\mathcal{P}$ and $\mathcal{T}$ are individually broken, their combination $\mathcal{PT}$ remains preserved. This symmetry structure plays a crucial role in determining the allowed NGM conductivity.

A minimal tight-binding Hamiltonian describing CuMnAs in momentum space can be written as~\cite{Kamal_Das_2023_PRB, Watanabe2021}
%%%%%%%%%%-------------------------------------------------------------
\begin{align}
H^{\rm CuMnAs}(\mathbf{k}) =
\begin{pmatrix}
\epsilon_0(\mathbf{k}) + \mathbf{h}_A(\mathbf{k})\cdot\boldsymbol{\sigma} & V_{AB}(\mathbf{k}) \\
V_{AB}(\mathbf{k}) & \epsilon_0(\mathbf{k}) + \mathbf{h}_B(\mathbf{k})\cdot\boldsymbol{\sigma}
\end{pmatrix},
\end{align}
%%%%%%%%%%-------------------------------------------------------------
where $\epsilon_0(\mathbf{k}) = -t(\cos k_x + \cos k_y)$, $t$ describes the intra-sublattice hopping and $V_{AB}(\mathbf{k}) = -2\tilde{t}\cos(k_x/2)\cos(k_y/2)$ , $\tilde{t}$  accounts for inter-sublattice hopping. The sublattice-dependent spin-orbit and exchange fields satisfy $\mathbf{h}_B(\mathbf{k}) = -\mathbf{h}_A(\mathbf{k})$, ensuring $\mathcal{PT}$ symmetry. The explicit form of $\mathbf{h}_A(\mathbf{k})$ is given by $\mathbf{h}_A(\mathbf{k}) =
\big(h_x^{\rm AFM} - \alpha_R \sin k_y + \alpha_D \sin k_y,\; h_y^{\rm AFM} + \alpha_R \sin k_x + \alpha_D \sin k_x,\;
h_z^{\rm AFM} \big)$, where $\alpha_R$ and $\alpha_D$ denote Rashba and Dresselhaus spin-orbit couplings, respectively, and $\mathbf{h}^{\rm AFM}$ represents the staggered exchange field. 
%%%%%%%%%%---------------------------------------------------------------
\begin{figure}[ht!]
\centering
\includegraphics[width=\columnwidth]{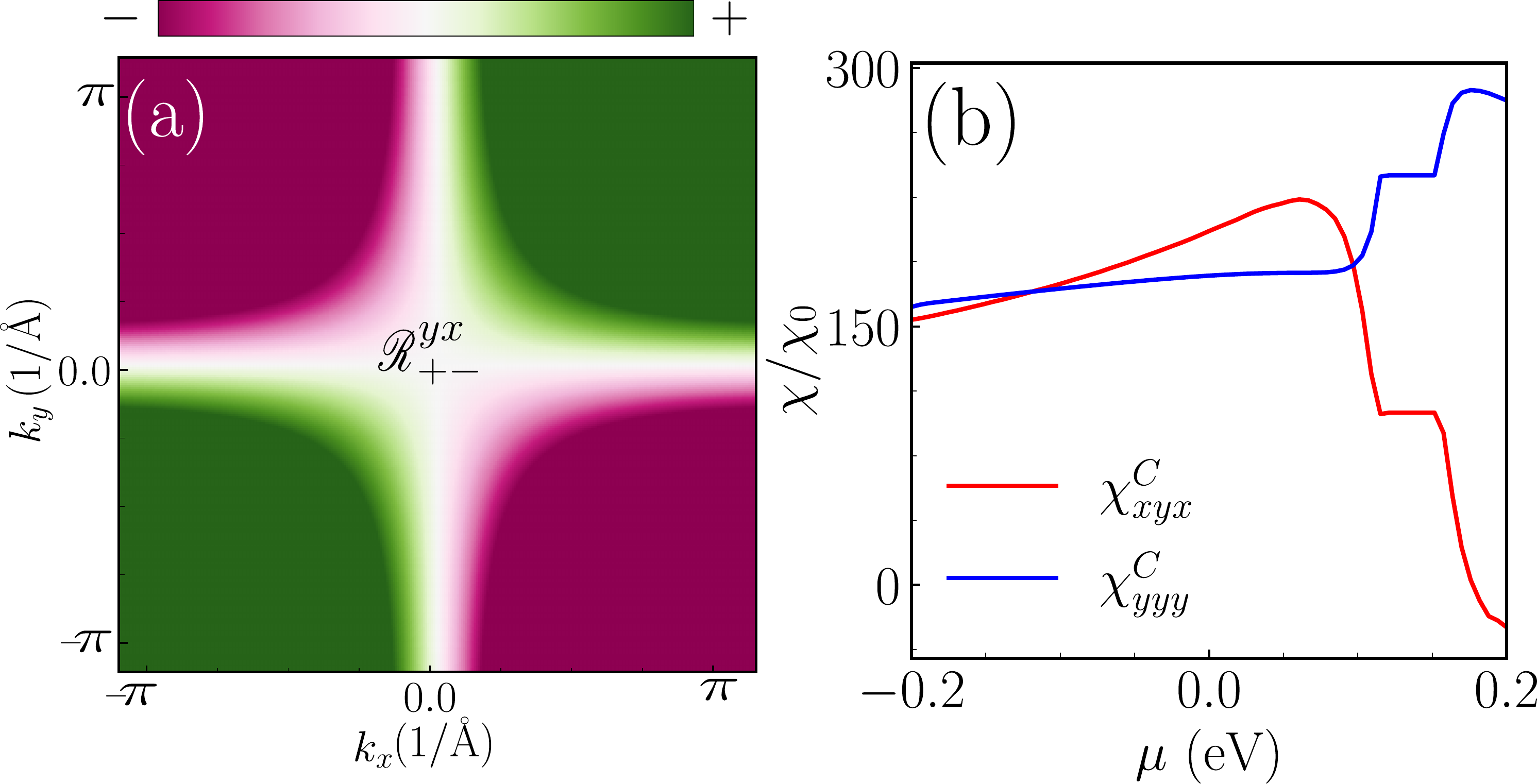}
\caption{(a) The off-diagonal component of spin-rotation quantum metric shows a quadrupolar character.
(b)The chemical potential ($\mu$) dependence of the some of the nonvanishing components of the conduction NGM conductivity is presented for a CuMnAs system.
The parameters used are $t_0 = 1~\text{eV}$, $\tilde{t} = 0.08~\text{eV}$, $h_A = [0.85, 0 ,0]~\text{eV}$, $t = 0.3~\text{eV}$, $\alpha_R = 0.08~\text{eV}$, $\alpha_D = 0.0~\text{eV}$, and $\chi_0 = \left(\frac{g\mu_B}{2}\right)^2 \,\mathrm{A\,m^{-1}\,T^{-2}}$.}
\label{Fig:CuMnAs}
\end{figure}
%%%%%%%%%%---------------------------------------------------------------
Under $\mathcal{PT}$ symmetry, the band energies satisfy $\epsilon_n(\mathbf{k})=\epsilon_n(-\mathbf{k})$. In such systems, the band-resolved SRBC $\Lambda^{ab}_{mp}$ vanishes identically, and consequently $\tilde{\mathscr{L}}^{abc}_{mp}$ is also strictly zero due to the absence of the Zeeman Berry curvature. As a result, the displacement current contribution to the NGM conductivity response vanishes identically. In contrast, the spin-rotation quantum metric SRQM $\mathscr{R}^{ab}_{mp}$ is odd under $\mathcal{PT}$ symmetry, while ${\mathscr{L}}^{abc}_{mp}$ is even. This symmetry structure allows both diagonal and off-diagonal components of the conduction current to be finite. This behavior stands in sharp contrast to $\mathcal{T}$-symmetric systems, where the conduction current vanishes and only the displacement current contributes to the nonlinear response.

Using the various components of the QGT, we evaluate the NGM conductivity and find that only the conduction current survives as depicted in Fig.~\ref{Fig:CuMnAs}, consistent with the symmetry analysis. We identify three independent components of the conduction NGM conductivity tensor, $\chi_{xxy}$, $\chi_{yyy}$, and $\chi_{yxx}$. Interestingly, the diagonal components of different NGM conductivity satisfy $\chi_{xxy}^{C,d} = \chi_{xyx}^{C,d}$ and for off-diagonal component $\chi_{yyy}^{C,od} = \chi_{yxx}^{C,od}$. This behavior can be understood from the symmetry properties of the underlying QGTs. The diagonal component $\chi^{C,d}$ is governed by SRQM $\Lambda^{ab}_{mp}$, which is symmetric under the exchange of indices $b$ and $c$, ensuring that $\chi_{xxy}^{C,d} = \chi_{xyx}^{C,d}$.

\section{Summary and conclusions}
In this work, we develop a microscopic quantum-kinetic framework to elucidate how the Zeeman and spin–rotation quantum geometric tensors (ZQGT and SRQGT) govern NGM transport in two-dimensional systems. Within a density-matrix formalism, we derive general expressions for second-order gyrotropic charge currents, incorporating both conduction and displacement contributions. We demonstrate that the microscopic structure of the nonlinear gyrotropic magnetic current tensor exhibits a clear separation between diagonal and off-diagonal components. For the
diagonal sector, the conduction contribution is dictated by the spin–rotation quantum metric, while the displacement response is governed by the spin–rotation Berry curvature. In contrast, off-diagonal sector, the conduction response is controlled by the Zeeman symplectic connection, whereas the displacement response originates from the Zeeman metric connection. Our analysis reveals that the SRQGT—absent in linear response—naturally emerges as the fundamental geometric quantity underlying NGM conductivity.

To demonstrate these effects, we analyze four representative systems with different symmetries: (i) two-dimensional massless Dirac system, (ii) surface states of a 3D topological insulator (TI) with hexagonal warping, (iii) tilted massive Dirac system, and (iv) $\mathcal{PT}$-symmetric CuMnAs system. These models capture distinct symmetry-breaking mechanisms in two-dimensional materials. For a massless Dirac system, the simultaneous presence of both mirror ($\mathcal{M}_x$ and $\mathcal{M}_y$) and time-reversal symmetries suppresses all NGM responses. In contrast, the presence of hexagonal warping in Dirac system breaks the $\mathcal{M}_y$ mirror symmetry while preserving $\mathcal{M}_x$ mirror symmetry and time-reversal symmetry activating certain components of the displacement-current channel, while the conduction-current contribution remains symmetry forbidden. Further, upon introducing a tilt and mass terms, all the mirror, time-reversal, and particle--hole symmetries are broken, thereby enabling both conduction- and displacement-current contributions to the NGM response. Finally, in the $\mathcal{PT}$-symmetric CuMnAs system, where time-reversal and inversion symmetries are explicitly broken but their combined $\mathcal{PT}$ symmetry is preserved, only the conduction-current contribution to the NGM response is allowed.

These findings show how distinct components of the ZQGT and SRQGT selectively govern nonlinear gyrotropic channels, offering symmetry-based design principles for engineering materials with tailored nonlinear magnetic responses. Perturbative magnetic-field corrections to the Berry curvature, neglected here, would only renormalize the intrinsic gyrotropic magnetic current without altering the symmetry classification of the geometric contributions~\cite{Wang_2024,Wang_2024_PRL}. Overall, our study establishes the ZQGT and SRQGT as key ingredients governing nonlinear hyrotropic magnetotransport responses that remain finite even when conventional Berry-curvature-driven effects vanish, providing an experimentally accessible probe of hidden spin-resolved quantum geometry in multiband systems.

\section{Acknowledgments}

N.~C. acknowledges the Council of Scientific and Industrial Research (CSIR), Government of India, for providing the JRF fellowship. S.~K.~G. acknowledges financial support from Anusandhan National Research Foundation (ANRF) erstwhile Science and Engineering Research Board (SERB), Government of India via the Startup Research Grant: SRG/2023/000934. S.~N. also acknowledges financial support from Anusandhan National Research Foundation (ANRF), Government of India via the Prime Minister's Early Career Research Grant: ANRF/ECRG/2024/005947/PMS.

\bibliography{Zeeman_NL}
\end{document}